\documentclass{jpsj2}
\usepackage{bm}

\title{Charge and Spin Currents Generated by Dynamical Spins}

\author{Akihito \textsc{Takeuchi}$^{1}$\thanks{E-mail address: atake@phys.metro-u.ac.jp} and Gen \textsc{Tatara}$^{1,2}$}

\inst{
$^{1}$Graduate School of Science and Engineering, Tokyo Metropolitan University, Hachioji, Tokyo 192-0397, Japan
\\
$^{2}$PRESTO, JST, 4-1-8 Honcho Kawaguchi, Saitama 332-0012, Japan
}

\abst{
We demonstrate theoretically that a charge current and a spin current are generated by spin dynamics in the presence of spin-orbit interaction in the perturbative regime.
We consider a general spin-orbit interaction including the spatially inhomogeneous case.
Spin current due to spin damping is identified as one origin of generated charge current, but other contributions exist, 
such as the one due to an induced conservative field and the one arising from the inhomogeneity of spin-orbit interaction.
}

\kword{
spintronics, 
inverse spin Hall effect,
spin current,
spin Hall effect, 
current generation,
spin battery,
Keldysh formalism
}

\begin{document}

\maketitle

\section{Introduction}
Spin Hall effect \cite{Hirsch,Zhang,Murakami,Sinova,Kato,Wunderlich} is one of the most interesting phenomena in spintronics,
which enables the control of magnetic properties by purely electrical means.
The idea is to induce a spin current in a transverse direction to an applied electric field by using spin-orbit interaction.
As an inverse effect, one can expect the conversion of spin current into charge current or electric voltage by use of spin-orbit interaction.
This effect, namely the inverse spin Hall effect, was proposed by Saitoh \textit{et al}. \cite{Saitoh}
and indeed observed experimentally in metallic systems \cite{Saitoh,Valenzuela,Kimura} and in semiconductor (GaAs) \cite{Zhao}.
One should note, however, that detection of spin Hall effect has so far been done by observing magnetization as a result of flow of spin current,
and not the spin current itself.
In the inverse effect, similarly, the electric voltage is measured as a response to an time-dependent external field
which drives magnetization dynamics.

The inverse of spin Hall effect was theoretically pointed out by Zhang and Niu \cite{ZhangNiu} and Hankiewicz \textit{et al}. \cite{Hankiewicz},
where they discussed a transverse charge current by a gradient of a spin-dependent chemical potential
(they called this effect the reciprocal spin Hall effect).
The spin-dependent chemical potential was argued to be related by an optical method.
In a junction of ferromagnet attached to normal metal, generation of electric voltage by applying an alternating magnetic field
observed by Costache \textit{et al}. \cite{Costache}.
Theoretical explanation of dc voltage was done by Wang \textit{et al}. \cite{Wang}
as due to the spin accumulation at the interface arising from a backflow of pumped spin current into the ferromagnet.
Voltage generation from spin dynamics was predicted by Stern \cite{Stern} in a slightly different context of Faraday's law for a fictitious field of Berry's phase.
The application to magnetic domain wall was done by Barnes and Maekawa \cite{BarnesMaekawa} and Duine \cite{Duine}.
In the Berry's phase mechanism, spin-orbit interaction is not essential but contributes as correction \cite{Duine}.

Direct relation between the pumped charge current and the magnetization dynamics was investigated recently by Ohe \textit{et al}. \cite{Ohe}.
They considered a disordered two-dimensional electron gas with the Rashba spin-orbit interaction and interacting with dynamical magnetization.
The charge current was calculated perturbatively.
Diffusive electron motion as represented by diffusion pole
(proportional to $1/q^2$ for small momentum transfer, $q$)
was taken account of since it leads to logarithmical long-range correlation in two-dimensions.
Considering a case of uniform Rashba system, they found that pumped charge current had a contribution proportional to $\langle{{\bm S}\times\dot{\bm S}}\rangle$,
where $\bm S$ is a local spin and $\langle{\cdots}\rangle$ here denotes average taking account the diffusive motion of electrons.
The quantity $\langle{{\bm S}\times\dot{\bm S}}\rangle$ represents a spin damping and is related phenomenologically to a spin current across the interface in the case of junctions
\cite{Tserkovnyak02,Tserkovnyak05}.
The result thus supports the idea of inverse spin Hall effect, where charge current is converted from the spin current.
Qualitatively this contribution to the current was, however, found and the inverse spin Hall mechanism turned out to be too naive.
Ohe \textit{et al}. also noted that the uniform Rashba system is peculiar, with many cancellations among the Feynman diagrams,
similarly to the peculiarity known in the spin Hall effect. \cite{Inoue}

In this paper we extend the study by Ohe \textit{et al}. \cite{Ohe} to the cases of general spin-orbit interaction,
including the case of inhomogeneous spin-orbit interaction.
The result can thus be applied to a finite Rashba system attached to leads.
It turns out that such inhomogeneity also contributes to a current with different symmetry proportional to $\langle{\dot{\bm S}}\rangle$.
We consider a three-dimensional case and so do not take account of diffusion ladders which
give only small contributions unlike in a two-dimensional case considered in ref.~\citen{Ohe}.

\section{System}
\begin{figure}[bt]
\begin{center}
\includegraphics[scale=0.8]{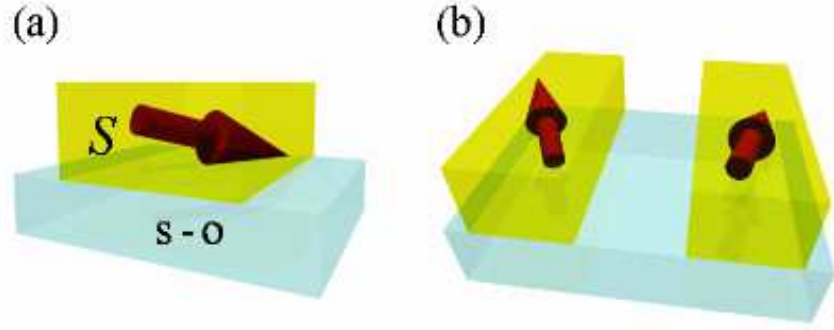}
\caption{(Color online).
Two typical systems with spin-orbit interaction and local spins.
The arrow is a local spin ($S$) which may have spatial structure and dynamics, 
and a bottom layer describes material with spin-orbit interaction (s-o).
}
\label{FIGsystem}
\end{center}
\end{figure}

We consider an electron system with spin-orbit interaction and exchange interaction with local spins.
The local spins can have arbitrary structure and thus
we can discuss various systems, such as those with one or two ferromagnets attached to a nonmagnet as shown in Fig.~\ref{FIGsystem}.
We consider a disordered system which would be the case of most experiments in metallic systems.
The total Hamiltonian is $H(t) = H_{0}+H_{\rm ex}(t)+H_{\rm so}+H_{\rm imp}$,
where 
\begin{subequations}\label{Hamiltonian}
\begin{gather}
H_{0} = -\frac{\hbar^2}{2m}\sum_{\bm x}\psi^\dagger_{\bm x}{\bm \nabla}^2\psi_{\bm x},
\\
H_{\rm ex}(t) = -J_{\rm ex}\sum_{\bm x}\psi^\dagger_{\bm x}\left[{\bm S}_{\bm x}(t)\cdot{\bm \sigma}\right]\psi_{\bm x},
\\
H_{\rm so} = -{\rm i}\sum_{\bm x}\psi^\dagger_{\bm x}\left\{\left[\left({\bm \nabla} U_{\bm x}\right)\times{\bm \nabla}\right]\cdot{\bm \sigma}\right\}\psi_{\bm x},
\\
H_{\rm imp} = u\sum_{i=1}^{n_{\rm i}}\psi^\dagger_{{\bm r}_i}\psi_{{\bm r}_i}.
\end{gather}
\end{subequations}
Here the annihilation (and creation) operator of conduction electrons in coordinate space is $\psi_{\bm x}$ (and $\psi^\dagger_{\bm x}$).
The first term describes free electron.
The second term $H_{\rm ex}$ denotes the exchange interaction,
where $J_{\rm ex}$ is a strength of the exchange coupling,
${\bm S}_{\bm x}(t)$ represents the local spins which can have any  spatial and slow temporal structure,
and $\bm \sigma$ represent Pauli matrices.
The spin-orbit interaction is represented by $H_{\rm so}$,
where $U_{\bm x}$ is a scalar potential (including a factor $\hbar^2/4m^2c^2$).
The last term $H_{\rm imp}$ is the spin-independent impurity scattering
which gives rise an elastic electron lifetime $\tau \equiv (2\pi N_{\rm e}n_{\rm i}u^2/\hbar V)^{-1}$,
where $u$ is a strength of the impurity scattering,
$n_{\rm i}$ is a number of impurities,
$N_{\rm e}$ is the electron's density of states at Fermi energy,
and $V$ is system volume.

\section{Charge Current}
The electron velocity operator is defined as $\hat{\bm v} = \frac{\rm i}{\hbar}[H,{\bm x}]$ ($[A,B]$ represents the commutator $AB-BA$),
which reads $\hat{v}_\mu = -\frac{{\rm i}\hbar}{m}\frac{\partial}{\partial x_\mu}+\frac{1}{\hbar}\epsilon_{\mu\nu\eta}\frac{\partial U_{\bm x}}{\partial x^{\eta}}\sigma^\nu$.
The charge current density is defined as ${\bm j}_{\rm c}({\bm x},t) \equiv -e\langle{\psi^\dagger_{\bm x}(t)\frac{\overleftrightarrow{\bm v}}{2}\psi_{\bm x}(t)}\rangle$,
where $A^\dagger\overleftrightarrow{\bm v}B \equiv (\hat{\bm v}A)^\dagger B+A^\dagger(\hat{\bm v}B)$ and $\langle{\cdots}\rangle$ is the expectation value estimated by the total Hamiltonian $H$.
It is given by
\begin{equation}\label{jcdef}
j_{\rm c\mu}({\bm x},t) =
e{\rm Tr}\left\{\left[\frac{\hbar^2}{2m}\left(\frac{\partial}{\partial x}-\frac{\partial}{\partial x'}\right)_\mu
+{\rm i}\epsilon_{\mu\nu\eta}\frac{\partial U_{\bm x}}{\partial x^\eta}\sigma^\nu\right]G^<({\bm x},t;{\bm x}',t)\right\}\Bigg|_{{\bm x}'={\bm x}},
\end{equation}
where ${\rm Tr}\{{\cdots}\}$ represents trace over spin indices and $G^<({\bm x},t;{\bm x}',t)$ represents a lesser Green function defined as
$G^<({\bm x},t;{\bm x}',t') \equiv \frac{\rm i}{\hbar}\langle{\psi^\dagger_{{\bm x}'}(t')\psi_{\bm x}(t)}\rangle$.
This charge current satisfies the charge continuity equation,
\begin{equation}\label{ChargeContinuity}
\frac{\partial \rho_{\rm c}({\bm x},t)}{\partial t}+{\bm \nabla}\cdot{\bm j}_{\rm c}({\bm x},t) = 0,
\end{equation}
where $\rho_{\rm c}({\bm x},t)$ ($\equiv -e\langle{\psi^\dagger_{\bm x}(t)\psi_{\bm x}(t)}\rangle$) is the charge density.

Assuming a dirty case ($J_{\rm ex} \ll \hbar/\tau$ and $k_{\rm F} U \ll \hbar/\tau$, $k_{\rm F}$ being Fermi momentum),
we carry out a perturbation expansion to calculate the charge current.
We treat the exchange interaction to the second order and the spin-orbit interaction to the first order.
Path ordered Green function \cite{HaugJauho,RammerSmith} is defined as
$G({\bm x},t;{\bm x}',t') \equiv -\frac{\rm i}{\hbar}\langle{{\rm T}_{C}\{\psi_{\bm x}(t)\psi^\dagger_{{\bm x}'}(t')\}}\rangle$,
where ${\rm T}_{C}\{{\cdots}\}$ is a path ordering operator defined on Keldysh contour $C$.
This Green function satisfies the Dyson equation on complex contour,
\begin{multline}\label{DysonEq}
G({\bm x},t;{\bm x}',t') =
g_{{\bm x}-{\bm x}'}(t-t')
-J_{\rm ex}\sum_{\bm X}\int_C{dt_{\rm ex}}g_{{\bm x}-{\bm X}}(t-t_{\rm ex})\left[{\bm S}_{\bm X}(t_{\rm ex})\cdot{\bm \sigma}\right]G({\bm X},t_{\rm ex};{\bm x}',t')
\\
-{\rm i}\sum_{\bm R}\int_C{dt_{\rm so}}g_{{\bm x}-{\bm R}}(t-t_{\rm so})\left\{\left[\left({\bm \nabla}_{\bm R} U_{\bm R}\right)\times{\bm \nabla}_{\bm R}\right]\cdot{\bm \sigma}\right\}G({\bm R},t_{\rm so};{\bm x}',t'),
\end{multline}
where $g_{\bm x}(t)$ denotes free Green function defined as
$g_{{\bm x}-{\bm x}'}(t-t') \equiv -\frac{\rm i}{\hbar}\langle{{\rm T}_{C}\{\psi_{\bm x}(t)\psi^\dagger_{{\bm x}'}(t')\}}\rangle_0$,
where $\langle{\cdots}\rangle_0$ is the expectation value estimated by free Hamiltonian $H_{0}$ and averaged over impurity scatterings.
Dyson equation is solved by iteration.
If non-interacting, the charge current in $\mu$-direction is simply proportional to $\langle{k_\mu}\rangle$ in momentum space.
This contribution vanishes, for a system is spatial symmetry (we assume this throughout this paper).
The charge current first order either in the exchange interaction or the spin-orbit interaction also vanishes since ${\rm Tr}\{{\sigma}\}=0$.
Therefore the charge current only arises if exchange and spin-orbit interactions couple.

\subsection{First order in $J_{\rm ex}$}
\begin{figure}[bt]
\begin{center}
\includegraphics[scale=0.8]{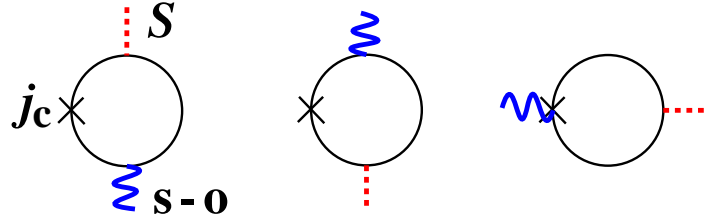}
\caption{(Color online).
Diagrammatic representations of the charge current at the first order in $J_{\rm ex}$.
Dotted lines and wavy lines denote the exchange interaction with local spin ($S$) and the spin-orbit interaction (s-o), respectively.}
\label{FIGjc1}
\end{center}
\end{figure}

By use of eqs.~(\ref{jcdef}) and (\ref{DysonEq}), contribution from the left diagram in Fig.~\ref{FIGjc1} is given by
\begin{multline}
j_{\rm c\mu}^{\rm (Fig.\ref{FIGjc1}-left)}({\bm x},t) =
\frac{{\rm i}e\hbar^2J_{\rm ex}}{m}\left(\frac{\partial}{\partial x}-\frac{\partial}{\partial x'}\right)_\mu\sum_{{\bm X},{\bm R}}\bigg[\int_C{dt_{1}}\int_C{dt_{2}}
{\rm Tr}\Big\{
g_{{\bm x}-{\bm X}}(t-t_{1})\left[{\bm S}_{\bm X}(t_1)\cdot{\bm \sigma}\right]
\\ \times 
g_{{\bm X}-{\bm R}}(t_{1}-t_{2})\left\{\left[\left({\bm \nabla}_{\bm R} U_{\bm R}\right)\times{\bm \nabla}_{\bm R}\right]\cdot{\bm \sigma}\right\}g_{{\bm R}-{\bm x}'}(t_{2}-t)
\Big\}\bigg]^<_{{\bm x}'={\bm x}},
\end{multline}
where $<$ denotes taking lesser component.
Using ${\rm Tr}\{{\sigma^\alpha\sigma^\beta}\} = 2\delta_{\alpha\beta}$ and calculating a lesser component,
we obtain
\begin{multline}
j_{\rm c\mu}^{\rm (Fig.\ref{FIGjc1}-left)}({\bm x},t) =
\frac{{\rm i}e\hbar^2J_{\rm ex}}{m}\left(\frac{\partial}{\partial x}-\frac{\partial}{\partial x'}\right)_\mu\epsilon_{\gamma\nu\eta}
\sum_{{\bm X},{\bm R}}\sum_{\omega,\Omega}{\rm e}^{{\rm i}\Omega t}\frac{\partial U_{\bm R}}{\partial R^\eta}S^\nu_{{\bm X},\Omega}
\\
\times\bigg\{
\left[f(\omega+\Omega)-f(\omega)\right]g^{\rm r}_{{\bm x}-{\bm X},\omega}g^{\rm a}_{{\bm X}-{\bm R},\omega+\Omega}\left(\frac{\partial}{\partial R^\gamma}g^{\rm a}_{{\bm R}-{\bm x}',\omega+\Omega}\right)
\\
+f(\omega)g^{\rm a}_{{\bm x}-{\bm X},\omega}g^{\rm a}_{{\bm X}-{\bm R},\omega+\Omega}\left(\frac{\partial}{\partial R^\gamma}g^{\rm a}_{{\bm R}-{\bm x}',\omega+\Omega}\right)
\\
-f(\omega+\Omega)g^{\rm r}_{{\bm x}-{\bm X},\omega}g^{\rm r}_{{\bm X}-{\bm R},\omega+\Omega}\left(\frac{\partial}{\partial R^\gamma}g^{\rm r}_{{\bm R}-{\bm x}',\omega+\Omega}\right)
\bigg\}\bigg|_{{\bm x}'={\bm x}},
\end{multline}
where $g^{\rm r}_{{\bm x},\omega} (= (g^{\rm a}_{{\bm x},\omega})^*) = \frac{1}{V}\sum_{\bm k}{\rm e}^{{\rm i}{\bm k}\cdot{\bm x}}g^{\rm r}_{{\bm k},\omega}$
($g^{\rm r}_{{\bm k},\omega} = (g^{\rm a}_{{\bm k},\omega})^* = (\hbar\omega-\varepsilon_{\bm k}+\varepsilon_{\rm F}+\frac{{\rm i}\hbar}{2\tau})^{-1}$,
where $\varepsilon_{\bm k} \equiv \hbar^2{\bm k}^2/2m$ and $\varepsilon_{\rm F}$ is Fermi energy),
$f(\omega)$ is the Fermi distribution function which is given as $f(\omega) \equiv \theta(-\omega)$ at zero temperature ($\theta(\omega)$ is a step function),
and ${\bm S}_{{\bm X},\Omega}$ is Fourier transform of ${\bm S}_{\bm X}(t)$.
Assuming that local spins vary slowly ($\Omega \ll \tau^{-1}$),
we obtain
\begin{equation}
j_{\rm c\mu}^{\rm (Fig.\ref{FIGjc1}-left)}({\bm x},t) \simeq
\frac{e\hbar^2J_{\rm ex}}{2\pi m}\left(\frac{\partial}{\partial x'}-\frac{\partial}{\partial x}\right)_{\mu}\epsilon_{\gamma\nu\eta}
\sum_{{\bm X},{\bm R}}\frac{\partial U_{\bm R}}{\partial R^\eta}\dot{S}^\nu_{\bm X}(t)
g^{\rm r}_{{\bm x}-{\bm X}}g^{\rm a}_{{\bm X}-{\bm R}}\left(\frac{\partial}{\partial R^\gamma}g^{\rm a}_{{\bm R}-{\bm x}'}\right)\bigg|_{{\bm x}'={\bm x}},
\end{equation}
where $g_{\bm x} \equiv g_{{\bm x},\omega=0}$.
We neglected terms containing only $g^{\rm r}$'s or $g^{\rm a}$'s since they are higher order of $\hbar/\varepsilon_{\rm F}\tau \ll 1$ as compared with mixed terms.
Other contributions are similarly calculated and the whole contribution in Fig.~\ref{FIGjc1} is obtained as
\begin{multline}\label{jc1coord}
j_{\rm c\mu}^{\rm (Fig.\ref{FIGjc1})}({\bm x},t) =
\frac{eJ_{\rm ex}}{\pi}\epsilon_{\mu\nu\eta}\frac{\partial U_{\bm x}}{\partial x^\eta}
{\rm Re}\sum_{\bm X}\dot{S}^\nu_{\bm X}(t)g^{\rm r}_{{\bm x}-{\bm X}}g^{\rm a}_{{\bm X}-{\bm x}}
\\
+\frac{e\hbar^2J_{\rm ex}}{\pi m}\frac{\partial}{\partial x_\mu}\epsilon_{\gamma\nu\eta}
{\rm Re}\sum_{{\bm X},{\bm R}}\frac{\partial U_{\bm R}}{\partial R^\eta}\dot{S}^\nu_{\bm X}(t)
g^{\rm r}_{{\bm x}-{\bm X}}\left(\frac{\partial}{\partial R^\gamma}g^{\rm a}_{{\bm X}-{\bm R}}\right)g^{\rm a}_{{\bm R}-{\bm x}}
\\
-\frac{2e\hbar^2J_{\rm ex}}{\pi m}\epsilon_{\gamma\nu\eta}
{\rm Re}\sum_{{\bm X},{\bm R}}\frac{\partial U_{\bm R}}{\partial R^\eta}\dot{S}^\nu_{\bm X}(t)
g^{\rm r}_{{\bm x}-{\bm X}}\left(\frac{\partial}{\partial R^\gamma}g^{\rm a}_{{\bm X}-{\bm R}}\right)\left(\frac{\partial}{\partial x_\mu}g^{\rm a}_{{\bm R}-{\bm x}}\right).
\end{multline}
The first term corresponds to the right diagram in Fig.~\ref{FIGjc1}, i.e., correction of the current vertex due to the spin-orbit interaction.
In momentum space, eq.~(\ref{jc1coord}) reads
\begin{multline}\label{jc1moment}
j_{\rm c\mu}^{\rm (Fig.\ref{FIGjc1})}({\bm x},t) =
-\frac{{\rm i}2e\hbar^{2}J_{\rm ex}}{\pi mV}\epsilon_{\gamma\nu\eta}
\sum_{{\bm q},{\bm p}}{\rm e}^{-{\rm i}({\bm q}+{\bm p})\cdot{\bm x}}p^\eta U_{\bm p}\dot{S}_{\bm q}^\nu(t)
\\
\times{\rm Re}\sum_{\bm k}
\left[\left({\bm k}-\frac{{\bm q}}{2}\right)_\mu k^\gamma g^{\rm r}_{{\bm k}-{\bm q}}g^{\rm a}_{\bm k}\left(g^{\rm a}_{{\bm k}+{\bm p}}-g^{\rm a}_{\bm k}\right)
+\frac{p_\mu}{2}k^\gamma g^{\rm r}_{{\bm k}-{\bm q}}g^{\rm a}_{\bm k}g^{\rm a}_{{\bm k}+{\bm p}}\right],
\end{multline}
where ${\bm S}_{\bm q}(t) \equiv \frac{1}{V}\sum_{\bm X}{\rm e}^{{\rm i}{\bm q}\cdot{\bm X}}{\bm S}_{\bm X}(t)$ and $U_{\bm p} \equiv \frac{1}{V}\sum_{\bm R}{\rm e}^{{\rm i}{\bm p}\cdot{\bm R}}U_{\bm R}$.

\subsection{Second order in $J_{\rm ex}$}
\begin{figure}[bt]
\begin{center}
\includegraphics[scale=0.8]{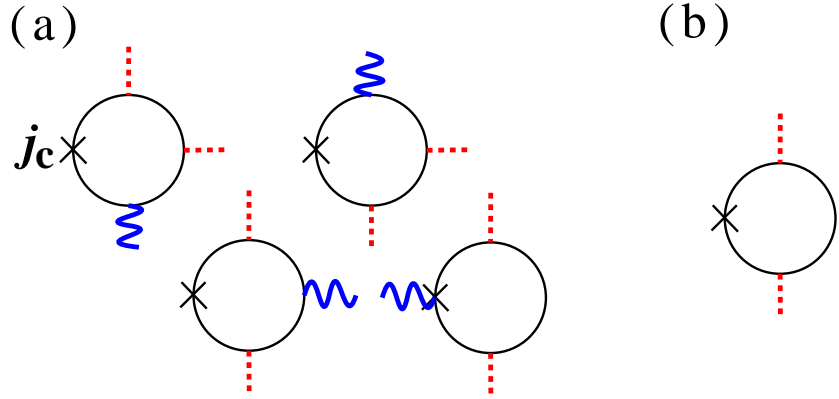}
\caption{(Color online).
Charge current at the second order in $J_{\rm ex}$.
(a) and (b) are in the presence and the absence of spin-orbit interaction, respectively.}
\label{FIGjc2}
\end{center}
\end{figure}
Contribution from the second order in $J_{\rm ex}$ shown in Fig.~\ref{FIGjc2}(a) is calculated similarly
by using ${\rm Tr}\{\sigma^\alpha\sigma^\beta\sigma^\gamma\} = {\rm i}2\epsilon_{\alpha\beta\gamma}$
as
\begin{multline}\label{jc2coord}
j_{\rm c\mu}^{\rm (Fig.\ref{FIGjc2}a)}({\bm x},t) =
\frac{2eJ_{\rm ex}^2}{\pi}\epsilon_{\mu\nu\eta}\frac{\partial U_{\bm x}}{\partial x^\eta}{\rm Im}\sum_{{\bm X}_1,{\bm X}_2}\left[{\bm S}_{{\bm X}_1}(t)\times\dot{\bm S}_{{\bm X}_2}(t)\right]^\nu
g^{\rm r}_{{\bm x}-{\bm X}_1}g^{\rm r}_{{\bm X}_1-{\bm X}_2}g^{\rm a}_{{\bm X}_2-{\bm x}}
\\
+\frac{e\hbar^2J_{\rm ex}^2}{\pi m}\epsilon_{\gamma\nu\eta}\frac{\partial}{\partial x_\mu}{\rm Im}\sum_{{\bm X}_1,{\bm X}_2,{\bm R}}\frac{\partial U_{\bm R}}{\partial R^\eta}\left[{\bm S}_{{\bm X}_1}(t)\times\dot{\bm S}_{{\bm X}_2}(t)\right]^\nu
\begin{cases}
+g^{\rm r}_{{\bm x}-{\bm R}}\left(\frac{\partial}{\partial R^\gamma}g^{\rm r}_{{\bm R}-{\bm X}_1}\right)g^{\rm r}_{{\bm X}_1-{\bm X}_2}g^{\rm a}_{{\bm X}_2-{\bm x}}
\\
+g^{\rm r}_{{\bm x}-{\bm X}_1}g^{\rm r}_{{\bm X}_1-{\bm X}_2}g^{\rm a}_{{\bm X}_2-{\bm R}}\left(\frac{\partial}{\partial R^\gamma}g^{\rm a}_{{\bm R}-{\bm x}}\right)
\\
+g^{\rm r}_{{\bm x}-{\bm X}_2}g^{\rm a}_{{\bm X}_2-{\bm R}}\left(\frac{\partial}{\partial R^\gamma}g^{\rm a}_{{\bm R}-{\bm X}_1}\right)g^{\rm a}_{{\bm X}_1-{\bm x}}
\end{cases}
\\
-\frac{2e\hbar^2J_{\rm ex}^2}{\pi m}\epsilon_{\gamma\nu\eta}{\rm Im}\sum_{{\bm X}_1,{\bm X}_2,{\bm R}}\frac{\partial U_{\bm R}}{\partial R^\eta}\left[{\bm S}_{{\bm X}_1}(t)\times\dot{\bm S}_{{\bm X}_2}(t)\right]^\nu
\begin{cases}
+\left(\frac{\partial}{\partial x_\mu}g^{\rm r}_{{\bm x}-{\bm R}}\right)\left(\frac{\partial}{\partial R^\gamma}g^{\rm r}_{{\bm R}-{\bm X}_1}\right)g^{\rm r}_{{\bm X}_1-{\bm X}_2}g^{\rm a}_{{\bm X}_2-{\bm x}}
\\
+\left(\frac{\partial}{\partial x_\mu}g^{\rm r}_{{\bm x}-{\bm X}_1}\right)g^{\rm r}_{{\bm X}_1-{\bm X}_2}g^{\rm a}_{{\bm X}_2-{\bm R}}\left(\frac{\partial}{\partial R^\gamma}g^{\rm a}_{{\bm R}-{\bm x}}\right)
\\
+\left(\frac{\partial}{\partial x_\mu}g^{\rm r}_{{\bm x}-{\bm X}_2}\right)g^{\rm a}_{{\bm X}_2-{\bm R}}\left(\frac{\partial}{\partial R^\gamma}g^{\rm a}_{{\bm R}-{\bm X}_1}\right)g^{\rm a}_{{\bm X}_1-{\bm x}}
\end{cases}.
\end{multline}
In momentum space, this is written as
\begin{multline}\label{jc2moment}
j_{\rm c\mu}^{\rm (Fig.\ref{FIGjc2}a)}({\bm x},t) =
-\frac{{\rm i}2e\hbar^2J_{\rm ex}^2}{\pi mV}\epsilon_{\gamma\nu\eta}
\sum_{{\bm q},{\bm Q},{\bm p}}{\rm e}^{-{\rm i}({\bm Q}+{\bm p})\cdot{\bm x}}p^\eta U_{\bm p}\left[{\bm S}_{{\bm Q}-{\bm q}}(t)\times\dot{\bm S}_{\bm q}(t)\right]^\nu
\\
\times{\rm Im}\sum_{\bm k}
\begin{cases}
+\left({\bm k}+\frac{{\bm Q}+{\bm p}}{2}\right)_{\mu}\left({\bm k}+{\bm q}\right)^\gamma g^{\rm r}_{\bm k}g^{\rm a}_{{\bm k}+{\bm q}}\left(g^{\rm a}_{{\bm k}+{\bm q}+{\bm p}}g^{\rm a}_{{\bm k}+{\bm Q}+{\bm p}}+g^{\rm a}_{{\bm k}+{\bm q}}g^{\rm a}_{{\bm k}+{\bm Q}}\right)
\\
-\left({\bm k}+\frac{{\bm Q}+{\bm p}}{2}\right)_{\mu}k^\gamma\left(g^{\rm r}_{{\bm k}-{\bm p}}-g^{\rm r}_{\bm k}\right)g^{\rm r}_{\bm k}g^{\rm a}_{{\bm k}+{\bm q}}g^{\rm a}_{{\bm k}+{\bm Q}}
\\
-\left({\bm k}+\frac{{\bm Q}+{\bm p}}{2}\right)_{\mu}\left({\bm k}+{\bm Q}\right)^\gamma g^{\rm r}_{\bm k}g^{\rm a}_{{\bm k}+{\bm q}}g^{\rm a}_{{\bm k}+{\bm Q}}\left(g^{\rm a}_{{\bm k}+{\bm Q}+{\bm p}}-g^{\rm a}_{{\bm k}+{\bm Q}}\right)
\\
+p_\mu k^\gamma g^{\rm r}_{{\bm k}-{\bm p}}g^{\rm r}_{\bm k}g^{\rm a}_{{\bm k}+{\bm q}}g^{\rm a}_{{\bm k}+{\bm Q}}
\end{cases}.
\end{multline}

The charge current arises even without the spin-orbit interaction at the second order in $J_{\rm ex}$.
This contribution is shown in Fig.~\ref{FIGjc2}(b) and given as
\begin{multline}\label{jc2nocoord}
j_{\rm c\mu}^{\rm (Fig.\ref{FIGjc2}b)}({\bm x},t) =
\frac{e\hbar^2J_{\rm ex}^2}{\pi m}\frac{\partial}{\partial x_\mu}
{\rm Im}\sum_{{\bm X}_1,{\bm X}_2}\left[{\bm S}_{{\bm X}_1}(t)\cdot\dot{\bm S}_{{\bm X}_2}(t)\right]
g^{\rm r}_{{\bm x}-{\bm X}_1}g^{\rm r}_{{\bm X}_1-{\bm X}_2}g^{\rm a}_{{\bm X}_2-{\bm x}}
\\
-\frac{2e\hbar^2J_{\rm ex}^2}{\pi m}
{\rm Im}\sum_{{\bm X}_1,{\bm X}_2}\left[{\bm S}_{{\bm X}_1}(t)\cdot\dot{\bm S}_{{\bm X}_2}(t)\right]
g^{\rm r}_{{\bm x}-{\bm X}_1}g^{\rm r}_{{\bm X}_1-{\bm X}_2}\left(\frac{\partial}{\partial x_\mu}g^{\rm a}_{{\bm X}_2-{\bm x}}\right).
\end{multline}
This current is simply due to chemical potential shift by the exchange interaction.
It contains a scalar product, ${\bm S}\cdot\dot{\bm S}$, in contrast with the contribution of Fig.~\ref{FIGjc2}(a), 
and so is small if local spins are spatially slowly varying.
In momentum space, eq.~(\ref{jc2nocoord}) is written as
\begin{equation}\label{jc2nomoment}
j_{\rm c\mu}^{\rm (Fig.\ref{FIGjc2}b)}({\bm x},t) =
-\frac{{\rm i}2e\hbar^2J_{\rm ex}^2}{\pi mV}
\sum_{{\bm q},{\bm Q}}{\rm e}^{-{\rm i}{\bm Q}\cdot{\bm x}}\left[{\bm S}_{{\bm Q}-{\bm q}}(t)\cdot\dot{\bm S}_{\bm q}(t)\right]
{\rm Im}\sum_{\bm k}\left({\bm k}+\frac{\bm Q}{2}\right)_\mu g^{\rm r}_{\bm k}g^{\rm a}_{{\bm k}+{\bm q}}g^{\rm a}_{{\bm k}+{\bm Q}}.
\end{equation}
The total charge current is given as
$j_{\rm c} = j_{\rm c}^{\rm (Fig.\ref{FIGjc1})}+j_{\rm c}^{\rm (Fig.\ref{FIGjc2}a)}+j_{\rm c}^{\rm (Fig.\ref{FIGjc2}b)}$.
This current contains a component which originates from a conservative scalar potential (or electric field) created by the spin dynamics and the spin-orbit interaction, such as the second terms of eqs.~(\ref{jc1coord}) and (\ref{jc2coord}) and the first term of eq.~(\ref{jc2nocoord}).
This scalar potential (or electric field) is fictitious, which only acts if the charge current has spin degrees of freedom.
The existence of this conservative part is in contrast to the adiabatic case in ref.~\citen{BarnesMaekawa}.

The above results are quite similar to those in two-dimensions \cite{Ohe}.
The difference is that averaging over long-range diffusion motion in two-dimensions is replaced in three-dimensions by a short-ranged average within the scale of electron mean free path.

\section{Spin Current}
We consider the spin current arising from exchange interaction.
We neglect here spin-orbit interaction because we are interested in conversion of spin current into charge current by spin-orbit interaction.
Spin current density is defined as $j_{\rm s\mu}^\nu({\bm x},t) \equiv \frac{\hbar}{2}\langle{\psi^\dagger_{\bm x}(t)\frac{\{\hat{v}_\mu,\sigma^\nu\}}{2}\psi_{\bm x}(t)}\rangle$
($\{A,B\}$ represents the anticommutator $AB+BA$ and $\hat{v}_\mu = -\frac{{\rm i}\hbar}{m}\frac{\partial}{\partial x_\mu}$), i.e.,
\begin{equation}\label{jsdef}
j_{\rm s\mu}^\nu({\bm x},t) =
\frac{\hbar^3}{4m}\left(\frac{\partial}{\partial x'}-\frac{\partial}{\partial x}\right)_\mu{\rm Tr}\left\{{\sigma^\nu G^<({\bm x},t;{\bm x}',t)}\right\}\Bigg|_{{\bm x}'={\bm x}}.
\end{equation}
In contrast to the charge continuity equation (eq.~(\ref{ChargeContinuity})), this spin current satisfies the equation,
\begin{equation}\label{SpinContinuity}
\frac{\partial \rho_{\rm s}^\nu({\bm x},t)}{\partial t}+{\bm \nabla}\cdot{\bm j}_{\rm s}^\nu({\bm x},t) = \mathcal{T}_{\rm s}^\nu({\bm x},t),
\end{equation}
where $\rho_{\rm s}^\nu({\bm x},t)$ ($\equiv \frac{\hbar}{2}\langle{\psi^\dagger_{\bm x}(t)\sigma^\nu\psi_{\bm x}(t)}\rangle$) is the spin density and
$\mathcal{T}_{\rm s}^\nu({\bm x},t)$ ($\equiv \frac{\rm i}{2}\langle{\psi^\dagger_{\bm x}(t)[H_{\rm ex},\sigma^\nu]\psi_{\bm x}(t)}\rangle$) is the torque density (or spin source)
due to exchange interaction.
Spin is thus not conserved.
(Definition of spin current has ambiguity when spin-orbit interaction is taken account \cite{SunXie,Shi},
but the continuity equation, eq.~(\ref{SpinContinuity}), is always satisfied with $\mathcal{T}_{\rm s}$ redefined properly.)

\begin{figure}[bt]
\begin{center}
\includegraphics[scale=0.8]{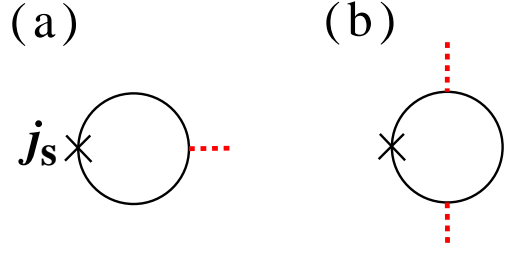}
\caption{(Color online).
Diagrams of the spin current.
(a) and (b) denote contributions from the first and the second order in $J_{\rm ex}$, respectively.}
\label{FIGjs}
\end{center}
\end{figure}

Diagrammatic representations of the spin current at the first and the second order in $J_{\rm ex}$ are shown in Fig.~\ref{FIGjs}.
These contributions are given by
\begin{equation}\label{js1coord}
j_{\rm s\mu}^{\rm \nu(Fig.\ref{FIGjs}a)}({\bm x},t) =
\frac{\hbar^3J_{\rm ex}}{2\pi m}{\rm Im}\sum_{\bm X}\dot{S}_{\bm X}^\nu(t)g^{\rm r}_{{\bm x}-{\bm X}}\left(\frac{\partial}{\partial x_\mu}g^{\rm a}_{{\bm X}-{\bm x}}\right),
\end{equation}
\begin{multline}\label{js2coord}
j_{\rm s\mu}^{\rm \nu(Fig.\ref{FIGjs}b)}({\bm x},t) =
\frac{\hbar^3J_{\rm ex}^2}{2\pi m}\frac{\partial}{\partial x_\mu}{\rm Re}\sum_{{\bm X}_1,{\bm X}_2}\left[{\bm S}_{{\bm X}_1}(t)\times\dot{\bm S}_{{\bm X}_2}(t)\right]^\nu
g^{\rm r}_{{\bm x}-{\bm X}_1}g^{\rm r}_{{\bm X}_1-{\bm X}_2}g^{\rm a}_{{\bm X}_2-{\bm x}}
\\
-\frac{\hbar^3J_{\rm ex}^2}{\pi m}{\rm Re}\sum_{{\bm X}_1,{\bm X}_2}\left[{\bm S}_{{\bm X}_1}(t)\times\dot{\bm S}_{{\bm X}_2}(t)\right]^\nu
g^{\rm r}_{{\bm x}-{\bm X}_1}g^{\rm r}_{{\bm X}_1-{\bm X}_2}\left(\frac{\partial}{\partial x_\mu}g^{\rm a}_{{\bm X}_2-{\bm x}}\right).
\end{multline}
They are written in momentum space as
\begin{equation}\label{js1moment}
j_{\rm s\mu}^{\rm \nu(Fig.\ref{FIGjs}a)}({\bm x},t) =
-\frac{{\rm i}\hbar^3J_{\rm ex}}{2\pi mV}
\sum_{\bm q}{\rm e}^{-{\rm i}{\bm q}\cdot{\bm x}}\dot{S}_{\bm q}^\nu(t)
{\rm Im}\sum_{\bm k}k_\mu g^{\rm r}_{{\bm k}-\frac{\bm q}{2}}g^{\rm a}_{{\bm k}+\frac{\bm q}{2}},
\end{equation}
\begin{equation}\label{js2moment}
j_{\rm s\mu}^{\rm \nu(Fig.\ref{FIGjs}b)}({\bm x},t) =
-\frac{{\rm i}\hbar^3J_{\rm ex}^2}{\pi mV}
\sum_{{\bm q},{\bm Q}}{\rm e}^{-{\rm i}{\bm Q}\cdot{\bm x}}\left[{\bm S}_{{\bm Q}-{\bm q}}(t)\times\dot{\bm S}_{\bm q}(t)\right]^\nu
{\rm Re}\sum_{\bm k}\left({\bm k}+\frac{\bm Q}{2}\right)_\mu g^{\rm r}_{\bm k}g^{\rm a}_{{\bm k}+{\bm q}}g^{\rm a}_{{\bm k}+{\bm Q}}.
\end{equation}
The total spin current contains terms like $\langle{\dot{\bm S}_{{\bm X}_1}}\rangle$ and $\langle{{\bm S}_{{\bm X}_1}\times\dot{\bm S}_{{\bm X}_2}}\rangle$, where
$\langle{\cdots}\rangle$ denotes average over position ${\bm X}_1$ and ${\bm X}_2$ within the length scale of electron mean free path.
Comparing these expressions with those of charge currents (eqs.~(\ref{jc1coord}) and (\ref{jc2coord})),
we see that they are related, suggesting the inverse spin Hall effect.
In fact, the first term in eq.~(\ref{jc1coord}) (eq.~(\ref{jc2coord})) arises from $\langle{\dot{\bm S}}\rangle$ ($\langle{{\bm S}\times\dot{\bm S}}\rangle$),
similarly to the spin current of eq.~(\ref{js1coord}) (eq.~(\ref{js2coord})).
These charge currents have extra factor of ${\bm \nabla} U_{\bm x}$ which justifies the idea of inverse spin Hall effect,
conversion of spin current into charge current by spin-orbit field, ${\bm E}_{\rm so}\equiv -{\bm \nabla} U_{\bm x}$.
However, other (second and third) terms of eqs.~(\ref{jc1coord}) and (\ref{jc2coord}) do not have direct correspondence to spin current.

\section{Smooth Spin Structures}
Let us first consider a case $U_{\bm p}$ is spatially smooth, i.e., $p \ll \ell^{-1}$, where $\ell$ is the electron mean free path.
We obtain the charge currents as
\begin{equation}
j_{\rm c\mu}^{\rm (Fig.\ref{FIGjc1})}({\bm x},t) =
\frac{{\rm i}e\hbar^2J_{\rm ex}}{2\pi mV}\epsilon_{\gamma\nu\eta}\frac{\partial^2 U_{\bm x}}{\partial x^\eta \partial x^\lambda}
\sum_{\bm q}{\rm e}^{-{\rm i}{\bm q}\cdot{\bm x}}\dot{S}_{\bm q}^\nu(t)
{\rm Re}\sum_{\bm k}\left(\delta_{\mu\lambda}k^\gamma-\delta_{\mu\gamma}k^\lambda\right)g^{\rm r}_{{\bm k}-{\bm q}}(g^{\rm a}_{\bm k})^2
+\mathcal{O}(p^3),
\end{equation}
\begin{multline}
j_{\rm c\mu}^{\rm (Fig.\ref{FIGjc2}a)}({\bm x},t) =
\frac{4e\hbar^2J_{\rm ex}^2}{\pi mV}\epsilon_{\gamma\nu\eta}\frac{\partial U_{\bm x}}{\partial x^\eta}
\sum_{{\bm q},{\bm Q}}{\rm e}^{-{\rm i}{\bm Q}\cdot{\bm x}}\left[{\bm S}_{{\bm Q}-{\bm q}}(t)\times\dot{\bm S}_{\bm q}(t)\right]^\nu
\\
\times{\rm Im}\sum_{\bm k}\left({\bm k}+\frac{\bm Q}{2}\right)_\mu\left({\bm k}+{\bm q}\right)^\gamma g^{\rm r}_{\bm k}(g^{\rm a}_{{\bm k}+{\bm q}})^2g^{\rm a}_{{\bm k}+{\bm Q}}
\\
+\frac{{\rm i}e\hbar^4J_{\rm ex}^2}{\pi m^2V}\epsilon_{\gamma\nu\eta}\frac{\partial^2 U_{\bm x}}{\partial x^\eta \partial x^\lambda}
\sum_{{\bm q},{\bm Q}}{\rm e}^{-{\rm i}{\bm Q}\cdot{\bm x}}\left[{\bm S}_{{\bm Q}-{\bm q}}(t)\times\dot{\bm S}_{\bm q}(t)\right]^\nu
\\
\times{\rm Im}\sum_{\bm k}
\begin{cases}
-\left({\bm k}+\frac{\bm Q}{2}\right)_\mu\left[k^\gamma\left({\bm k}+{\bm q}\right)^\lambda+\left({\bm k}+{\bm q}\right)^\gamma k^\lambda\right](g^{\rm r}_{\bm k})^2(g^{\rm a}_{{\bm k}+{\bm q}})^2g^{\rm a}_{{\bm k}+{\bm Q}}
\\
-\left({\bm k}+\frac{\bm Q}{2}\right)_\mu\left[k^\gamma Q^\lambda-k^\lambda Q^\gamma\right](g^{\rm r}_{\bm k})^2g^{\rm a}_{{\bm k}+{\bm q}}(g^{\rm a}_{{\bm k}+{\bm Q}})^2
\\
+\left({\bm k}+\frac{\bm Q}{2}\right)_\mu\left[\left({\bm k}+{\bm q}\right)^\gamma\left({\bm k}+{\bm Q}\right)^\lambda+\left({\bm k}+{\bm q}\right)^\gamma\left({\bm k}+{\bm Q}\right)^\lambda\right]g^{\rm r}_{\bm k}(g^{\rm a}_{{\bm k}+{\bm q}})^2(g^{\rm a}_{{\bm k}+{\bm Q}})^2
\end{cases}
\\
+\mathcal{O}(p^3).
\end{multline}
We also assume local spins vary slowly in space, and explore behavior of the currents in detail.
Carrying out $q$ and $Q$ expansion assuming as $q,Q \ll \ell^{-1}$, the currents are given as
\begin{multline}
j_{\rm c\mu}({\bm x},t) \sim
\frac{{\rm i}e\hbar^2J_{\rm ex}}{3\pi mV}\epsilon_{\gamma\nu\eta}\frac{\partial^2 U_{\bm x}}{\partial x^\eta \partial x^\lambda}
\sum_{\bm q}\left(\delta_{\mu\gamma}q^\lambda-\delta_{\mu\lambda}q^\gamma\right){\rm e}^{-{\rm i}{\bm q}\cdot{\bm x}}\dot{S}_{\bm q}^\nu(t)
{\rm Re}\sum_{\bm k}\varepsilon_{\bm k}(g^{\rm r}_{\bm k})^2(g^{\rm a}_{\bm k})^2
\\
+\frac{8eJ_{\rm ex}^2}{3\pi V}\epsilon_{\mu\nu\eta}\frac{\partial U_{\bm x}}{\partial x^\eta}
\sum_{{\bm q},{\bm Q}}{\rm e}^{-{\rm i}{\bm Q}\cdot{\bm x}}\left[{\bm S}_{{\bm Q}-{\bm q}}(t)\times\dot{\bm S}_{\bm q}(t)\right]^\nu
{\rm Im}\sum_{\bm k}\varepsilon_{\bm k}g^{\rm r}_{\bm k}(g^{\rm a}_{\bm k})^3
\\
+\frac{{\rm i}2e\hbar^2J_{\rm ex}^2}{3\pi mV}\epsilon_{\gamma\nu\eta}\frac{\partial^2 U_{\bm x}}{\partial x^\eta \partial x^\lambda}
\sum_{{\bm q},{\bm Q}}{\rm e}^{-{\rm i}{\bm Q}\cdot{\bm x}}\left[{\bm S}_{{\bm Q}-{\bm q}}(t)\times\dot{\bm S}_{\bm q}(t)\right]^\nu
{\rm Im}A_{{\bm q},{\bm Q}}^{\mu\gamma\lambda}
\\
-\frac{{\rm i}2e\hbar^2J_{\rm ex}^2}{\pi mV}
\sum_{{\bm q},{\bm Q}}{\rm e}^{-{\rm i}{\bm Q}\cdot{\bm x}}\left[{\bm S}_{{\bm Q}-{\bm q}}(t)\cdot\dot{\bm S}_{\bm q}(t)\right]
{\rm Im}B_{{\bm q},{\bm Q}}^\mu,
\end{multline}
\begin{multline}
j_{\rm s\mu}^\nu({\bm x},t) \sim
-\frac{{\rm i}\hbar^3J_{\rm ex}}{3\pi mV}
\sum_{\bm q}q_\mu{\rm e}^{-{\rm i}{\bm q}\cdot{\bm x}}\dot{S}_{\bm q}^\nu(t)
{\rm Im}\sum_{\bm k}\varepsilon_{\bm k}g^{\rm r}_{\bm k}(g^{\rm a}_{\bm k})^2
\\
-\frac{{\rm i}\hbar^3J_{\rm ex}^2}{\pi mV}
\sum_{{\bm q},{\bm Q}}{\rm e}^{-{\rm i}{\bm Q}\cdot{\bm x}}\left[{\bm S}_{{\bm Q}-{\bm q}}(t)\times\dot{\bm S}_{\bm q}(t)\right]^\nu
{\rm Re}B_{{\bm q},{\bm Q}}^\mu,
\end{multline}
where
\begin{multline}
A_{{\bm q},{\bm Q}}^{\mu\gamma\lambda} \equiv
\left[\delta_{\mu\gamma}\left({\bm q}+{\bm Q}\right)^\lambda+\delta_{\mu\lambda}\left({\bm q}+{\bm Q}\right)^\gamma\right]
\sum_{\bm k}\varepsilon_{\bm k}g^{\rm r}_{\bm k}(g^{\rm a}_{\bm k})^4
\\
-\left[\delta_{\mu\gamma}\left({\bm q}+{\bm Q}\right)^\lambda+\delta_{\mu\lambda}\left({\bm q}-{\bm Q}\right)^\gamma\right]
\sum_{\bm k}\varepsilon_{\bm k}(g^{\rm r}_{\bm k})^2(g^{\rm a}_{\bm k})^3
\\
+\frac{8}{3}\left[\delta_{\mu\gamma}\left({\bm q}+{\bm Q}\right)^\lambda+\delta_{\mu\lambda}\left({\bm q}+{\bm Q}\right)^\gamma\right]
\sum_{\bm k}\varepsilon_{\bm k}^2g^{\rm r}_{\bm k}(g^{\rm a}_{\bm k})^5
\\
-\frac{4}{3}\left[\delta_{\mu\gamma}\left(2{\bm q}+{\bm Q}\right)^\lambda+\delta_{\mu\lambda}\left(2{\bm q}+{\bm Q}\right)^\gamma\right]
\sum_{\bm k}\varepsilon_{\bm k}^2(g^{\rm r}_{\bm k})^2(g^{\rm a}_{\bm k})^4, \notag
\end{multline}
\begin{equation}
B_{{\bm q},{\bm Q}}^\mu \equiv
\sum_{\bm k}\left[\frac{Q_\mu}{2}g^{\rm r}_{\bm k}(g^{\rm a}_{\bm k})^2+\frac{2\left({\bm q}+{\bm Q}\right)_\mu}{3}\varepsilon_{\bm k}g^{\rm r}_{\bm k}(g^{\rm a}_{\bm k})^3\right]. \notag
\end{equation}
Here we used $\langle{k^\alpha k^\beta}\rangle = \delta_{\alpha\beta}{\bm k}^2/3$ where the average is over direction.
Sums over $k$ are carried out as 
\begin{gather}
\sum_{\bm k}\varepsilon_{\bm k}(g^{\rm r}_{\bm k})^2(g^{\rm a}_{\bm k})^2 \sim
\frac{4\pi N_{\rm e}\varepsilon_{\rm F}\tau^3}{\hbar^3}, \notag
\\
\sum_{\bm k}\varepsilon_{\bm k}g^{\rm r}_{\bm k}(g^{\rm a}_{\bm k})^3 \sim
-\frac{{\rm i}3\pi N_{\rm e}\tau^2}{2\hbar^2}-\frac{2\pi N_{\rm e}\varepsilon_{\rm F}\tau^3}{\hbar^3}, \notag
\\
\sum_{\bm k}\varepsilon_{\bm k}g^{\rm r}_{\bm k}(g^{\rm a}_{\bm k})^4 \sim
-\frac{{\rm i}2\pi N_{\rm e}\varepsilon_{\rm F}\tau^4}{\hbar^4}, \notag
\\
\sum_{\bm k}\varepsilon_{\bm k}(g^{\rm r}_{\bm k})^2(g^{\rm a}_{\bm k})^3 \sim
\frac{{\rm i}6\pi N_{\rm e}\varepsilon_{\rm F}\tau^4}{\hbar^4}, \notag
\\
\sum_{\bm k}\varepsilon_{\bm k}^2g^{\rm r}_{\bm k}(g^{\rm a}_{\bm k})^5 \sim
\frac{{\rm i}5\pi N_{\rm e}\varepsilon_{\rm F}\tau^4}{2\hbar^4}, \notag
\\
\sum_{\bm k}\varepsilon_{\bm k}^2(g^{\rm r}_{\bm k})^2(g^{\rm a}_{\bm k})^4 \sim
-\frac{{\rm i}5\pi N_{\rm e}\varepsilon_{\rm F}\tau^4}{\hbar^4}, \notag
\\
\sum_{\bm k}g^{\rm r}_{\bm k}(g^{\rm a}_{\bm k})^2 \sim
\frac{{\rm i}2\pi N_{\rm e}\tau^2}{\hbar^2}, \notag
\\
\sum_{\bm k}\varepsilon_{\bm k}g^{\rm r}_{\bm k}(g^{\rm a}_{\bm k})^{2} \sim
-\frac{3\pi N_{\rm e}\tau}{2\hbar}+\frac{{\rm i}2\pi N_{\rm e}\varepsilon_{\rm F}\tau^2}{\hbar^2}, \notag
\end{gather}
where $N_{\rm e}$ is the density of states given as $mVk_{\rm F}/2\pi^2\hbar^2$.
We finally obtain
\begin{multline}\label{jc}
j_{\rm c\mu}({\bm x},t) =
\frac{4eN_{\rm e}J_{\rm ex}\varepsilon_{\rm F}\tau^3}{3\hbar mV}
\left[
\epsilon_{\gamma\nu\eta}\frac{\partial^2 U_{\bm x}}{\partial x_\mu \partial x^\eta}\frac{\partial \dot{S}_{\bm x}^\nu(t)}{\partial x^\gamma}
-\epsilon_{\mu\nu\eta}\frac{\partial^2 U_{\bm x}}{\partial x^\eta \partial x^\lambda}\frac{\partial \dot{S}_{\bm x}^\nu(t)}{\partial x^\lambda}
\right]
\\
-\frac{4eN_{\rm e}J_{\rm ex}^2\tau^2}{\hbar^2V}\epsilon_{\mu\nu\eta}\frac{\partial U_{\bm x}}{\partial x^\eta}
\left[{\bm S}_{\bm x}(t)\times\dot{\bm S}_{\bm x}(t)\right]^\nu
\\
-\frac{8eN_{\rm e}J_{\rm ex}^2\varepsilon_{\rm F}\tau^4}{9\hbar^2mV}
\begin{cases}
+\epsilon_{\mu\nu\eta}\frac{\partial^2 U_{\bm x}}{\partial x^\eta \partial x^\lambda}
\left\{
4\frac{\partial}{\partial x^\lambda}\left[{\bm S}_{\bm x}(t)\times\dot{\bm S}_{\bm x}(t)\right]^\nu+9\left[{\bm S}_{\bm x}(t)\times\frac{\partial \dot{\bm S}_{\bm x}(t)}{\partial x^\lambda}\right]^\nu
\right\}
\\
+\epsilon_{\gamma\nu\eta}\frac{\partial^2 U_{\bm x}}{\partial x_\mu \partial x^\eta}
\left\{
13\frac{\partial}{\partial x^\gamma}\left[{\bm S}_{\bm x}(t)\times\dot{\bm S}_{\bm x}(t)\right]^\nu+9\left[{\bm S}_{\bm x}(t)\times\frac{\partial \dot{\bm S}_{\bm x}(t)}{\partial x^\gamma}\right]^\nu
\right\}
\end{cases}
\\
-\frac{2eN_{\rm e}J_{\rm ex}^2\tau^2}{mV}
\left[{\bm S}_{\bm x}(t)\cdot\frac{\partial \dot{\bm S}_{\bm x}(t)}{\partial x_\mu}\right],
\end{multline}
\begin{multline}\label{js}
j_{\rm s\mu}^\nu({\bm x},t) =
\frac{2\hbar N_{\rm e}J_{\rm ex}\varepsilon_{\rm F}\tau^2}{3mV}
\frac{\partial \dot{S}_{\bm x}^\nu(t)}{\partial x_\mu}
\\
-\frac{4N_{\rm e}J_{\rm ex}^2\varepsilon_{\rm F}\tau^3}{3mV}
\left\{
\frac{\partial}{\partial x_\mu}\left[{\bm S}_{\bm x}(t)\times\dot{\bm S}_{\bm x}(t)\right]^\nu
+\left[{\bm S}_{\bm x}(t)\times\frac{\partial \dot{\bm S}_{\bm x}(t)}{\partial x_\mu}\right]^\nu
\right\}.
\end{multline}
We see in eq.~(\ref{jc}) that
charge current driven by uniform spin-orbit field (${\bm E}_{\rm so}=-{\bm \nabla} U_{\bm x}$) is perpendicular to spin damping, ${\bm S}\times\dot{\bm S}$,
while discontinuity of ${\bm E}_{\rm so}$ induces new $\dot{\bm S}$ components.

In a case that local spins are spatially uniform, ${\bm S}_{\bm x}(t) = {\bm S}(t)$,
the spin current totally vanishes and the charge current only arises as
\begin{equation}
j_{\rm c\mu}^{\rm uniform}({\bm x},t) =
-\frac{4eN_{\rm e}J_{\rm ex}^2\tau^2}{\hbar^2V}\epsilon_{\mu\nu\eta}\frac{\partial U_{\bm x}}{\partial x^\eta}
\left[{\bm S}(t)\times\dot{\bm S}(t)\right]^\nu.
\end{equation}

\subsection{Rashba system}
We apply above result to the case of Rashba spin-orbit interaction.
Rashba spin-orbit interaction can be realized in various three-dimensional systems without inversion symmetry \cite{Fujimoto,Bauer}.
Assuming uniform spin-orbit field in $z$-direction, $-{\bm \nabla} U_{\bm x} = E_{\rm so}{\bf e}_z$, we obtain the charge current as
\begin{equation}
j_{\rm c\mu}^{\rm Rashba}({\bm x},t) =
\frac{4eN_{\rm e}E_{\rm so}J_{\rm ex}^2\tau^2}{\hbar^2V}\epsilon_{\mu\nu z}
\left[{\bm S}_{\bm x}(t)\times\dot{\bm S}_{\bm x}(t)\right]^\nu
-\frac{2eN_{\rm e}J_{\rm ex}^2\tau^2}{mV}
\left[{\bm S}_{\bm x}(t)\cdot\frac{\partial \dot{\bm S}_{\bm x}(t)}{\partial x_\mu}\right].
\end{equation}
It is interesting to note that in the case of inhomogeneous spin-orbit field, an additional current arises as
\begin{multline}
j_{\rm c\mu}^{\rm add}({\bm x},t) =
\frac{4eN_{\rm e}J_{\rm ex}\varepsilon_{\rm F}\tau^{3}}{3\hbar mV}
\left[
\epsilon_{\mu\nu z}\frac{\partial E_{\rm so}}{\partial x^\lambda}\frac{\partial \dot{S}_{\bm x}^\nu(t)}{\partial x^\lambda}
-\epsilon_{\gamma\nu z}\frac{\partial E_{\rm so}}{\partial x_\mu}\frac{\partial \dot{S}_{\bm x}^\nu(t)}{\partial x^\gamma}
\right]
\\
+\frac{8eN_{\rm e}J_{\rm ex}^2\varepsilon_{\rm F}\tau^4}{9\hbar^2mV}
\begin{cases}
+\epsilon_{\mu\nu z}\frac{\partial E_{\rm so}}{\partial x^\lambda}
\left\{
4\frac{\partial}{\partial x^\lambda}\left[{\bm S}_{\bm x}(t)\times\dot{\bm S}_{\bm x}(t)\right]^\nu+9\left[{\bm S}_{\bm x}(t)\times\frac{\partial \dot{\bm S}_{\bm x}(t)}{\partial x^\lambda}\right]^\nu
\right\}
\\
+\epsilon_{\gamma\nu z}\frac{\partial E_{\rm so}}{\partial x_\mu}
\left\{
13\frac{\partial}{\partial x^\gamma}\left[{\bm S}_{\bm x}(t)\times\dot{\bm S}_{\bm x}(t)\right]^\nu+9\left[{\bm S}_{\bm x}(t)\times\frac{\partial \dot{\bm S}_{\bm x}(t)}{\partial x^\gamma}\right]^\nu
\right\}
\end{cases}.
\end{multline}
This additional current appears, for instance,
at the interface between Rashba region and leads when normal leads are attached.

\section{Equilibrium Components}
In the above discussion on pumped currents, terms like $(g^{\rm r})^n$ and $(g^{\rm a})^n$ were negligibly small.
These contributions in fact correspond to the equilibrium contributions which exist even if local spins have no dynamics, ${\bm S}_{\bm x}(t) = {\bm S}_{\bm x}$.
The currents are then given by
\begin{multline}
j_{\rm c\mu}^{\rm (eq)}({\bm x}) =
4eJ_{\rm ex}\epsilon_{\mu\nu\eta}\frac{\partial U_{\bm x}}{\partial x^\eta}{\rm Im}\sum_{\bm X}S^\nu_{\bm X}\sum_\omega f(\omega)
g^{\rm a}_{{\bm x}-{\bm X},\omega}g^{\rm a}_{{\bm X}-{\bm x},\omega}
\\
+\frac{4e\hbar^2J_{\rm ex}}{m}\epsilon_{\gamma\nu\eta}\frac{\partial}{\partial x_\mu}{\rm Im}\sum_{{\bm X},{\bm R}}\frac{\partial U_{\bm R}}{\partial R^\eta}S^\nu_{\bm X}\sum_\omega f(\omega)
g^{\rm a}_{{\bm x}-{\bm X},\omega}\left(\frac{\partial}{\partial R^\gamma}g^{\rm a}_{{\bm X}-{\bm R},\omega}\right)g^{\rm a}_{{\bm R}-{\bm x},\omega}
\\
-\frac{8e\hbar^2J_{\rm ex}}{m}\epsilon_{\gamma\nu\eta}{\rm Im}\sum_{{\bm X},{\bm R}}\frac{\partial U_{\bm R}}{\partial R^\eta}S^\nu_{\bm X}\sum_\omega f(\omega)
g^{\rm a}_{{\bm x}-{\bm X},\omega}\left(\frac{\partial}{\partial R^\gamma}g^{\rm a}_{{\bm X}-{\bm R},\omega}\right)\left(\frac{\partial}{\partial x_\mu}g^{\rm a}_{{\bm R}-{\bm x},\omega}\right),
\end{multline}
\begin{multline}
j_{\rm s\mu}^{\rm \nu (eq)}({\bm x}) =
\frac{\hbar^3J_{\rm ex}^2}{m}\frac{\partial}{\partial x_\mu}{\rm Im}\sum_{{\bm X}_1,{\bm X}_2}\left[{\bm S}_{{\bm X}_1}\times{\bm S}_{{\bm X}_2}\right]^\nu\sum_\omega f(\omega)
g^{\rm a}_{{\bm x}-{\bm X}_1,\omega}g^{\rm a}_{{\bm X}_1-{\bm X}_2,\omega}g^{\rm a}_{{\bm X}_2-{\bm x},\omega}
\\
-\frac{2\hbar^3J_{\rm ex}^2}{m}{\rm Im}\sum_{{\bm X}_1,{\bm X}_2}\left[{\bm S}_{{\bm X}_1}\times{\bm S}_{{\bm X}_2}\right]^\nu\sum_\omega f(\omega)
g^{\rm a}_{{\bm x}-{\bm X}_1,\omega}g^{\rm a}_{{\bm X}_1-{\bm X}_2,\omega}\left(\frac{\partial}{\partial x_\mu}g^{\rm a}_{{\bm X}_2-{\bm x},\omega}\right).
\end{multline}
The equilibrium charge current thus does not contain $\langle{{\bm S}_{{\bm X}_1}\times{\bm S}_{{\bm X}_2}}\rangle$ term ($j_{\rm c}^{\rm (eq)} \propto \langle{{\bm S}_{\bm X}}\rangle$),
while spin current does not have $\langle{{\bm S}_{\bm X}}\rangle$ term
($j_{\rm s}^{\rm (eq)} \propto \langle{{\bm S}_{{\bm X}_1}\times{\bm S}_{{\bm X}_2}}\rangle$).
In momentum space they are written as
\begin{multline}
j_{\rm c\mu}^{\rm (eq)}({\bm x}) =
-\frac{{\rm i}8e\hbar^2J_{\rm ex}}{mV}\epsilon_{\gamma\nu\eta}
\sum_{{\bm q},{\bm p}}{\rm e}^{-{\rm i}({\bm q}+{\bm p})\cdot{\bm x}}p^\eta U_{\bm p}S_{\bm q}^\nu\sum_\omega f(\omega)
\\
\times{\rm Im}\sum_{\bm k}\left[\left({\bm k}-\frac{\bm q}{2}\right)_{\mu}k^\gamma g^{\rm a}_{{\bm k}-{\bm q},\omega}g^{\rm a}_{{\bm k},\omega}(g^{\rm a}_{{\bm k}+{\bm p},\omega}-g^{\rm a}_{{\bm k},\omega})
+\frac{p_\mu}{2}k^\gamma g^{\rm a}_{{\bm k}-{\bm q},\omega}g^{\rm a}_{{\bm k},\omega}g^{\rm a}_{{\bm k}+{\bm p},\omega}\right],
\end{multline}
\begin{equation}
j_{\rm s\mu}^{\rm \nu (eq)}({\bm x}) =
-\frac{{\rm i}2\hbar^3J_{\rm ex}^2}{mV}
\sum_{{\bm q},{\bm Q}}{\rm e}^{-{\rm i}{\bm Q}\cdot{\bm x}}\left[{\bm S}_{{\bm Q}-{\bm q}}\times{\bm S}_{\bm q}\right]^\nu\sum_\omega f(\omega)
{\rm Im}\sum_{\bm k}\left({\bm k}+\frac{\bm Q}{2}\right)_\mu g^{\rm a}_{{\bm k},\omega}g^{\rm a}_{{\bm k}+{\bm q},\omega}g^{\rm a}_{{\bm k}+{\bm Q},\omega}.
\end{equation}
The equilibrium currents calculated above correspond to an initial response to local spins suddenly attached to the system.
In reality, as soon as the equilibrium charge and spin currents appear, the spin configuration starts to evolve and dynamical currents appear.

In a spatially smooth local spins and spin-orbit interaction,
the equilibrium components of the charge current contain at least three derivatives,
$({\bm \nabla} U_{\bm x})({\bm \nabla}^2{\bm S}_{\bm x})$ or $({\bm \nabla}^2U_{\bm x})({\bm \nabla}{\bm S}_{\bm x})$ or $({\bm \nabla}^3U_{\bm x}){\bm S}_{\bm x}$, and are negligibly small.
The equilibrium spin current, in contrast, is large and given as
\begin{equation}
j_{\rm s\mu}^{\rm \nu (eq)}({\bm x}) \sim
-\frac{\hbar^2N_{\rm e}J_{\rm ex}^2}{12mV\varepsilon_{\rm F}}
\left[{\bm S}_{\bm x}\times\frac{\partial {\bm S}_{\bm x}}{\partial x_\mu}\right]^\nu.
\end{equation}
This result indicates that the equilibrium component of spin current cannot be converted into charge one.
This is physically reasonable since equilibrium spin current can be dissipationless while charge current always has dissipation.
Dynamical spins are thus essential in pumping of charge current.

\section{Conclusion}
We have discussed theoretically that charge and spin currents are induced by spin dynamics in the presence of spin-orbit interaction.
The spin current pumped from dynamical spins was found to contain two contributions, $\langle{\dot{\bm S}}\rangle$ and $\langle{{\bm S}\times\dot{\bm S}}\rangle$,
consistent with phenomenological argument \cite{Saitoh,Tserkovnyak02}.
We found that pumped charge current has a component flowing perpendicular to both spin-orbit field (${\bm E}_{\rm so}$) and average spin polarization $\langle{{\bm S}\times\dot{\bm S}}\rangle$.
This contribution is thus explained as due to conversion of spin current ($\propto \langle{{\bm S}\times\dot{\bm S}}\rangle$) into charge current, i.e., the inverse spin Hall effect.
For a pumping mechanism by use of local spins, dynamical spins are essential because the static spin current is not converted into charge current.
The charge current has another component originating from a fictitious conservative field which only acts if the charge current has spin degrees of freedom.
This current is not associated directly with spin current.
We also found a novel charge current proportional to $\partial{\bm E}_{\rm so}$, arising from inhomogeneity of spin-orbit field.
This contribution would be essential in actual experiments on finite size spin-orbit system attached to leads.

\section*{Acknowledgments}
We are grateful to J.-I. Ohe and E. Saitoh for helpful discussions and comments.


\begin{thebibliography}{99}
\bibitem{Hirsch}
J.E. Hirsch:
Phys. Rev. Lett. \textbf{83} (1999) 1834.

\bibitem{Zhang}
S. Zhang:
Phys. Rev. Lett. \textbf{85} (2000) 393.

\bibitem{Murakami}
S. Murakami, N. Nagaosa, and S.C. Zhang:
Science \textbf{301} (2003) 1348.

\bibitem{Sinova}
J. Sinova, D. Culcer, Q. Niu, N.A. Sinitsyn, T. Jungwirth, and A.H. MacDonald:
Phys. Rev. Lett. \textbf{92} (2004) 126603.

\bibitem{Kato}
Y.K. Kato, R.C. Myers, A.C. Gossard, and D.D. Awschalom:
Science \textbf{306} (2004) 1910.

\bibitem{Wunderlich}
J. Wunderlich, B. Kaestner, J. Sinova, and T. Jungwirth:
Phy. Rev. Lett. \textbf{94} (2005) 047204.

\bibitem{Saitoh}
E. Saitoh, M. Ueda, H. Miyajima, and G. Tatara:
Appl. Phys. Lett. \textbf{88} (2006) 182509.

\bibitem{Valenzuela}
S.O. Valenzuela and M. Tinkham:
Nature \textbf{442} (2006) 176.

\bibitem{Kimura}
T. Kimura, Y. Otani, T. Sato, S. Takahashi, and S. Maekawa:
Phys. Rev. Lett. \textbf{98} (2007) 156601.

\bibitem{Zhao}
H. Zhao, E.J. Loren, H.M. van Driel, and A.L. Smirl:
Phys. Rev. Lett. \textbf{96} (2006) 246601.

\bibitem{ZhangNiu}
P. Zhang and Q. Niu:
cond-mat/0406436.

\bibitem{Hankiewicz}
E.M. Hankiewicz, J. Li, T. Jungwirth, Q. Niu, S.-Q. Shen, and J. Sinova:
Phys. Rev. B \textbf{72} (2005) 155305.

\bibitem{Costache}
M.V. Costache, M. Sladkov, S.M. Watts, C.H. van der Wal, and B.J. van Wees:
Phys. Rev. Lett. \textbf{97} (2006) 216603.

\bibitem{Wang}
X. Wang, G.E.W. Bauer, B.J. van Wees, A. Brataas, and Y. Tserkovnyak:
Phys. Rev. Lett. \textbf{97} (2006) 216602.

\bibitem{Stern}
A. Stern:
Phys. Rev. Lett. \textbf{68} (1992) 1022.

\bibitem{BarnesMaekawa}
S.E. Barnes and S. Maekawa:
Phys. Rev. Lett. \textbf{98} (2007) 246601.

\bibitem{Duine}
R.A. Duine:
Phys. Rev. B \textbf{77} (2008) 014409.

\bibitem{Ohe}
J.-I. Ohe, A. Takeuchi, and G. Tatara:
Phys. Rev. Lett. \textbf{99} (2007) 266603.

\bibitem{Tserkovnyak02}
Y. Tserkovnyak, A. Brataas, and G.E.W. Bauer:
Phys. Rev. Lett. \textbf{88} (2002) 117601.

\bibitem{Tserkovnyak05}
Y. Tserkovnyak, A. Brataas, G.E.W. Bauer, and B.I. Halperin:
Rev. Mod. Phys. \textbf{77} (2005) 1375.

\bibitem{Inoue}
J. Inoue, G.E.W. Bauer, and L.W. Molenkamp:
Phys. Rev. B \textbf{70} (2004) 041303(R).

\bibitem{HaugJauho}
H. Haug and A.-P. Jauho:
\textit{Quantum Kinetics in Transport and Optics of Semiconductors}
(Springer-Verlag, 1998).

\bibitem{RammerSmith}
J. Rammer and H. Smith:
Rev. Mod. Phys. \textbf{58} (1986) 323.

\bibitem{SunXie}
Q.-F. Sun and X.C. Xie:
Phys. Rev. B \textbf{72} (2005) 245305.

\bibitem{Shi}
J. Shi, P. Zhang, D. Xiao, and Q. Niu:
Phys. Rev. Lett. \textbf{96} (2006) 076604.

\bibitem{Fujimoto}
S. Fujimoto:
J. Phys. Soc. Jpn. \textbf{76} (2007) 051008.

\bibitem{Bauer}
E. Bauer, H. Kaldarar, A. Prokofiev, E. Royanian, A. Amato, J. Sereni, W. Br\"amer-Escamilla, and I. Bonalde:
J. Phys. Soc. Jpn. \textbf{76} (2007) 051009.
\end{thebibliography}
\end{document}